\documentclass[epj]{webofc}
\usepackage[varg]{txfonts}   
\wocname{EPJ Web of Conferences}
\woctitle{New Frontiers in Physics 2014}
\usepackage{graphicx,amsmath,amssymb,bm}

\newcommand{\exclude}[1]{}

\newcommand{\beq}{\begin{equation}}
\newcommand{\eeq}{\end{equation}}
\newcommand{\be}{\begin{eqnarray}}
\newcommand{\ee}{\end{eqnarray}}
   
\def\dd{ \,\mathrm{d} }

\def\+{\dagger}
 \def\la{\langle}
 \def\ra{\rangle}

 \def\la{\langle}
\def\ra{\rangle}
\def\<{\langle}
\def\>{\rangle}
\newcommand{\Lqcd}{\Lambda_{\mathrm{QCD}}}
\newcommand{\Lbar}{\Lambda_{\overline{\mathrm{QCD}}}}
\newcommand{\qcd}{{\overline{\mathrm{QCD}}}}
    
\begin{document}
\selectlanguage{english}
\title{The topological long range order in QCD.  Applications to heavy ion collisions  and cosmology. 
}
%
%

\author{Ariel R. Zhitnitsky\inst{1}\fnsep\thanks{\email{arz@phas.ubc.ca}}}
              
\institute{Department of Physics and Astronomy, University of British Columbia, Vancouver, B.C. V6T 1Z1, Canada}
\abstract{%
We argue   that the local violation of  P invariance in heavy ion collisions is a consequence of the long range topological order  which is inherent feature of strongly coupled QCD. A similar phenomenon is known to occur in   some    topologically ordered  condensed matter systems with a gap.    
 We also discuss possible cosmological applications of this long range order in strongly coupled gauge theories. In particular, we argue that the de Sitter behaviour   might be dynamically generated as a result of the long range order. In this framework the inflaton   is an auxiliary    field which effectively describes the dynamics of topological sectors in a gauge theory    in the expanding universe, rather than a new dynamical degree of freedom.}
\maketitle
\section{Introduction}
\label{intro}
Recently it has become clear that quantum anomalies   play  very important role in the macroscopic dynamics of relativistic fluids. Much of this progress is motivated by very interesting ongoing  experiments 
on local ${\cal{P}}$  and ${\cal{CP}}$ violation in QCD as studied  at RHIC  \cite{Voloshin:2004vk,Abelev:2009tx,Wang:2012qs} and, more recently, at the LHC   \cite{Selyuzhenkov:2012py,Abelev:2012pa,Voloshin:2012fv,Selyuzhenkov:2012mf}. It is likely that the observed asymmetry is due to   charge separation effect \cite{Kharzeev:2007tn} as a result of the chiral anomaly, see recent review \cite{Kharzeev:2013ffa} as  an introduction to the subject. 
  The ideas formulated in \cite{Kharzeev:2007tn}  were further developed  in follow up papers \cite{Kharzeev:2007jp,Fukushima:2008xe}  where the effect was coined as chiral magnetic effect (CME). 

We  shall not discuss a number of subtle questions  of the  CME in the present work by  referring  to a recent review \cite{Kharzeev:2013ffa}. Instead, we concentrate on a single  crucial element for CME to be operational. Namely, the key assumption of the proposal \cite{Kharzeev:2007tn,Kharzeev:2013ffa}  is that the region where  the so-called $\la\theta (\vec{x}, t)_{ind}\ra\neq 0$  should be much larger in size than the scale of  conventional QCD fluctuations which have  typical     correlation lengths of order $\sim \Lambda_{QCD}^{-1}$. The  $\theta (\vec{x}, t)_{ind}$ parameter enters the effective lagrangian as follows,  ${\cal L_{\theta}}=-\theta_{ind} q$ where $ q \equiv \frac{g^2}{64\pi^2} \epsilon_{\mu\nu\rho\sigma} G^{a\mu\nu} G^{a\rho\sigma}$ is the topological density operator, such that local ${\cal{P}}$  and ${\cal{CP}}$
  invariance of QCD is broken on the scales where correlated state with $\la\theta (\vec{x}, t)_{ind}\ra\neq 0$  is induced. One should expect a number of ${\cal{P}}$  and ${\cal{CP}}$ violating effects taking place in a relatively large  region where $\la\theta (\vec{x}, t)_{ind}\ra\neq 0$.  
  
  The first question to be  addressed in the present work is as follows: what is the physics behind of this long range order in the system with a gap $\sim \Lqcd$? As it is known, normally, if the system has a gap, conventional  correlation functions     decay exponentially fast with a typical   scale of order of $\Lqcd^{-1}$ rather than demonstrate a long range order which is normally associated with a massless particle.    The second question to be addressed in the present work  can be formulated as follows. Let us assume that the long range order is  indeed  present in the system.  Than,    how  one can   observe it?
    
  One should   comment here that the charge separation effect, CME, chiral separation effect  and many other  related phenomena 
 have been very  active area of research in recent years, see   review paper  \cite{Kharzeev:2013ffa} with large number of references on the original results. However, 
     the key ingredient of the entire framework,  the presence of the long range structure in the system,  has not been properly addressed in previous studies\footnote{In particular, 
     for the effective Lagrangian approach  advocated in the original proposal \cite{Kharzeev:2007tn} (as well as in  many followup papers) to be justified,     the corresponding slow parameter  $\theta (\vec{x}, t)_{ind} \neq 0$ may only fluctuate with frequencies which are much smaller than $\Lqcd$. Otherwise, the effective Lagrangian approach   breaks down. The presence of the new scale of the problem can be formally expressed as $|\partial_{\mu}\theta_{ind}| \sim R^{-1}\ll \Lqcd$, where parameter $R$ can be identified with the size of the system.    The problem on the long range order in this framework  can be formulated as follows: how one could understand  the emergence of the  new scale $\sim R$ in the system with the  gap $\sim \Lqcd$? The same question in terms of the axial chemical potential defined as   $   \mu_5\equiv \dot{\theta}_{ind}$ \cite{Fukushima:2008xe,Kharzeev:2013ffa} can be formulated as follows: why  $\mu_5$ is coherently correlated   on  a  large scale of order $\sim R$?    }, with very few exceptions.
     The goal of this work is to explain the emergence of this long range order  and present few profound observational consequences which might be related to this long range structure. In our presentation we follow the original results  on the subject \cite{Zhitnitsky:2013hs,Zhitnitsky:2012im,Zhitnitsky:2012ej,Zhitnitsky:2013pna,Zhitnitsky:2014aja}.
     
  Our presentation is organized as follows. We start  in section \ref{contact}  with definition of  the key 
 element  of our  study--the non-dispersive contact term  in topological susceptibility. The unique feature of this contact term is that  it  does   contribute  to  
    the  $\theta$ dependent    portion of  the vacuum energy $E_{\mathrm{vac}}(\theta)$. However, this term   can not be expressed in terms of  any physical propagating degrees of freedom, therefore, it has a non-dispersive nature.   Exactly this fundamentally new type of the vacuum energy is highly sensitive to the long distance physics, and will play the  key role in  our applications considered in section \ref{ions} and \ref{cosmology}. Specifically, in section \ref{contact} we shall argue that the long range sensitivity is related to the non-perturbative 
     dynamics of the  topologically  nontrivial  sectors in gauge theories \cite{Zhitnitsky:2013hs}. 
    In section \ref{ions} we apply our findings on long range order    to heavy ion collisions \cite{Zhitnitsky:2012im,Zhitnitsky:2012ej}, while in section \ref{cosmology} we consider some cosmological applications.   Specifically, we shall argue \cite{Zhitnitsky:2013pna,Zhitnitsky:2014aja} that the    de Sitter phase  may dynamically emerge  in strongly coupled gauge theories  as a result of   dynamics of the  topologically  nontrivial  sectors in expanding universe. 
If this effect is  confirmed by future analytical and numerical studies (which are presently underway), it may have profound consequences for our understanding of the inflationary phase in early inverse. It may also shed some light  on the  dynamics of the  dark energy we are witnessing now.

\section{Contact term, and the Long Range Order  in QCD}\label{contact}
 We   introduce  the topological susceptibility $\chi$
 which plays a crucial role in resolution of the   $U(1)_A$ problem in QCD ~\cite{witten,ven,vendiv} as follows\footnote{We use the Euclidean notations  where  path integral computations are normally performed.}
\be
\label{chi}
 \chi (\theta =0) =    \left. \frac{\partial^2E_{\mathrm{vac}}(\theta)}{\partial \theta^2} \right|_{\theta=0}= \lim_{k\rightarrow 0} \int \!\dd^4x e^{ikx} \la T\{q(x), q(0)\}\ra  ,
 \ee
where     $\theta$ parameter   enters the  Lagrangian  along with  topological density operator $q (x)$ and $E_{\mathrm{vac}}(\theta)$ is the $\theta$ dependent portion  of the  vacuum energy density  in QCD.   
It is important  that the topological susceptibility $\chi$  does not vanish in spite of the fact that $q= \partial_{\mu}K^{\mu}$ is total divergence. This feature is very different from any conventional correlation functions  which normally must  vanish  at zero momentum if the  corresponding operator  can be represented as 
total divergence.  Furthermore, any physical massive  state (meson, or glueball) gives a negative contribution to this 
diagonal correlation function
\be	\label{G}
  \chi_{\rm dispersive} \sim  \lim_{k\rightarrow 0} \int d^4x e^{ikx} \la T\{q(x), q(0)\}\ra  
  \sim 
    \lim_{k\rightarrow 0}  \sum_n \frac{\la  0 |q|n\ra \la n| q| 0\ra }{-k^2-m_n^2}\simeq -\sum_n\frac{|c_n|^2}{m_n^2} \leq 0,  
\ee
 where   $m_n$ is the mass of a physical $|n\ra$ state,  $k\rightarrow 0$  is  its momentum, and $\la 0| q| n\ra= c_n$ is its coupling to topological density operator $q (x)$.
 At the same time the resolution of the $U(1)_A$ problem requires a positive sign for the topological susceptibility (\ref{chi}), see the original reference~\cite{vendiv} for a thorough discussion, 
\be	\label{top1}
  \chi_{\rm non-dispersive}= \lim_{k\rightarrow 0} \int \!\dd^4x e^{ikx} \la T\{q(x), q(0)\}\ra > 0.~~~
\ee
Therefore, there must be a contact contribution to $\chi$, which is not related to any propagating  physical degrees of freedom,  and it must have the ``wrong" sign. The ``wrong" sign in this paper implies a sign 
  which is opposite to any contributions related to the  physical propagating degrees of freedom (\ref{G}). 
  The vacuum energy $E_{\mathrm{vac}}(\theta) $ which is expressed in terms of $\chi$ according to eq. (\ref{chi})   {\it can not} be formulated  in terms of any conventional propagating degrees of freedom as it has pure non-dispersive nature according to eqs. (\ref{G}, \ref{top1}).

    In the framework \cite{witten} the contact term with ``wrong" sign  has been simply postulated, while in refs.\cite{ven,vendiv} the Veneziano ghost (with a ``wrong" kinetic term) had been introduced into the theory to saturate the required property (\ref{top1}).   
  Furthermore, as we discuss below the  contact term has  the structure $\chi \sim \int d^4x \delta^4 (x)$.
  The significance of this structure is  that the gauge variant correlation function in momentum space
  \be
  \label{K}
   \lim_{k\rightarrow 0} \int d^4x e^{ikx} \la K_{\mu}(x) , K_{\nu}(0)\ra\sim   \frac{k_{\mu}k_{\nu}}{k^4}
   \ee 
  develops  a topologically protected  ``unphysical" pole which does not correspond to any propagating massless degrees of freedom, but nevertheless must be present in the system. Furthermore, the residue of this   pole has the ``wrong sign", which  precisely corresponds to the Veneziano ghost  contribution saturating the non-dispersive term  in gauge invariant correlation function (\ref{top1}),
   \be
  \label{K1}
   \< q({x}) q({0}) \> \sim  \la \partial_{\mu}K^{\mu}(x) , \partial_{\nu}K^{\nu}(0)\ra \sim \delta^4(x).
   \ee 
   We  should comment here, that the entire framework, including the singular behaviour of
  $ \< q({x}) q({0}) \>$   with the ``wrong sign",  has been  accepted by the community as a standard resolution of the $U(1)_A$ problem which  has been  well confirmed by numerous  lattice simulations in strongly  coupled regime, see e.g. recent papers \cite{Ilgenfritz:2007xu,Ilgenfritz:2008ia,Bruckmann:2011ve} 
  and many references therein.   Furthermore, it has been argued long ago in  \cite{Luscher:1978rn}
  that the gauge theories may exhibit the ``secret long range forces" (which is, in fact, the title of ref.\cite{Luscher:1978rn}) expressed in terms of the topologically protected massless pole in correlation function (\ref{K}). 
  
 While the the presence of the contact term (\ref{top1})  in the topological susceptibility in  QCD  is well 
 established fact, its  nature and origin  still remain  a mystery even today. Indeed, answers on the questions on   nature of the vacuum configurations  saturating  the topological susceptibility  (\ref{top1}) are  still unknown in strongly coupled QCD as the corresponding answers are hard to extract  from the numerical QCD lattice  simulations. 
 
 Fortunately, all relevant questions can be addressed  in a  weakly coupled gauge theory, the so-called ``deformed QCD" \cite{Yaffe:2008}. This model is a weakly coupled gauge theory due to some special deformations as constructed in \cite{Yaffe:2008}. However, this model   preserves all crucial ingredients of the original theory, such as confinement, the presence of the topological sectors, the $\theta$ dependence, etc.   which allow to perform  the analytical   computations   in theoretically controllable way. It  allows us to answer the crucial questions about the vacuum configurations   saturating the contact term in this model. Furthermore, it allows us to understand and illuminate  the  infra-red (IR) nature of the contact term, and demonstrate  its sensitivity to the large distance physics in spite of the fact that the system is gapped 
 as a result of confinement in this model. 
  
To be more specific, the topological susceptibility and the vacuum energy in the ``deformed QCD"   model 
 can be explicitly calculated, and  it is saturated by fractionally charged weakly interacting monopoles when semiclassical  computations are under complete theoretical control. The result  for $ |\theta|\leq \pi  $ is \cite{Thomas:2011ee,Zhitnitsky:2013hs}:
\be \label{YM}
  \chi_{YM}= \frac{\zeta}{NL} \int d^3 x \left[ \delta(\bold{x}) \right], ~~~
  E_{\rm vac}(\theta)=-\frac{N\zeta}{L} \cos \frac{\theta}{N}, ~~~  
   \la q(x)\ra \equiv -i\frac{\partial E_{\mathrm{vac}}(\theta)}{\partial \theta}=-i\frac{\xi}{L}\sin \frac{\theta}{N}.
   \ee
 In this formula  $N$ stands for number of colours, $\zeta$ is the monopole fugacity of the system, which is assumed to be parametrically small for semiclassical approximation to be justified, and parameter $L$ is the size of compact dimension in the Euclidean time direction.
  
The topological susceptibility  has the required  ``wrong sign" as this  contribution is not related to any physical propagating degrees of freedom, and it has a 
$\delta(\bold{x})$ function structure which implies the presence of the pole (\ref{K}). However, there are no any physical massless states in the system as  it  is gapped. This contact term (\ref{YM}) is described in terms of the tunnelling events between  different (but physically equivalent) topological sectors $|k\ra$ which always present in gauge   systems. 
The monopoles in this framework are not real particles, they are pseudo-particles which live in Euclidean space-time  and describe the physical tunnelling processes between different topological sectors $|k\ra$ and $| k+1 \ra$. The vacuum  energy of the system (\ref{YM}) should be interpreted  as number of tunnelling events per unit time $L$  per unit volume $V$
as explained in  \cite{Thomas:2011ee,Zhitnitsky:2013hs}. 
\exclude{
\be
\label{zeta}
  \left(\frac{  {\rm number ~of  ~tunnelling ~ events}}{VL}\right)=\frac{N\zeta}{L},~~~~~~~~~~ E_{\mathrm{vac}}
 =- \frac{N\zeta}{L}, ~~~~~
 \ee
where $\zeta$ is the monopole fugacity to be understood as a number of tunnelling events for a given type of monopole per unit time $L$. There are $N$ different types of monopoles which explains the normalization in eq.(\ref{zeta}). }
Precisely this interpretation reveals   the non-dispersive nature of this vacuum  energy which can not be attributed to any physical propagating degrees of freedom. It is quite obvious that the nature of this  energy is very different from conventional vacuum energy formulated in terms of any  dynamical propagating  field.
 
 Furthermore, the $\delta (\mathbf{x})$ function in  (\ref{YM}) should be understood as total divergence related to the IR  physics, rather than to ultraviolet (UV) behaviour  as explained in~\cite{Thomas:2011ee}
\be	\label{divergence}
 \chi_{YM}\, \sim   \int \delta (\mathbf{x})  d^3x =  \int   d^3x~
  \partial_{\mu}\left(\frac{x^{\mu}}{4\pi x^3}\right)=
    \oint_{S_2}    \dd\Sigma_{\mu}
 \left(\frac{x^{\mu}}{4\pi x^3}\right).
\ee
 In other words, the non-dispersive contact term with the ``wrong" sign (\ref{YM}) is {\it highly sensitive} to the boundary conditions and behaviour of the system at arbitrarily  large distances. Therefore, it is natural to expect that a variation of the boundary conditions or placement this  system into a time dependent background would change the topological susceptibility (\ref{YM})  and associated with it the vacuum energy $E_{\mathrm{vac}}$ despite of the fact that the system has a gap.  
 
 Our final remark in this section is as follows. We observe the long range sensitivity of the system in terms of topologically protected pole (\ref{K}), or what is the same, in terms of the surface integral (\ref{divergence})
 despite the presence of the gap in the system which itself   reflects  the property of the  confinement in this model. 
 One  can describe the same physics, including the behaviour in the IR region (\ref{K})    using the auxiliary topological non-propagating fields  as developed in ref. \cite{Zhitnitsky:2013hs}. In fact the corresponding construction \cite{Zhitnitsky:2013hs} in ``deformed QCD" model  sheds some lights on 
the nature of the Veneziano ghost \cite{ven,vendiv} which was introduced ad hoc in QCD long ago with the only purpose to saturate the correlation function (\ref{K}) with a ``wrong sign". One can explicitly see in this ``deformed QCD" model that  auxiliary topological non-propagating fields can be explicitly identified with the Veneziano ghost as these massless auxiliary  fields  exactly saturate the topological susceptibility without  violation  unitarity, casualty or any other fundamental properties of quantum field theory. This  representation  further illuminates the IR nature of the contact term
as the  auxiliary non-propagating topological fields are massless and indeed  develop a   pole at $k=0$ in eq. (\ref{K}). Furthermore, based on this construction, one can argue that QCD belongs to a topologically ordered phase, similar to many known examples in condensed matter physics, see  \cite{Zhitnitsky:2013hs} for the details.

\section{${\cal P}$ -odd fluctuations and Long Range Order in Heavy Ion Collisions}\label{ions}
In previous section we interpreted the vacuum  configurations which saturate the topological susceptibility (\ref{YM})  and associated with it the vacuum energy $E_{\mathrm{vac}}$ as  tunnelling processes which are happening all the time in Minkowski vacuum with no  interruptions.  These tunnelling processes in Minkowski vacuum (when no any external sources are present in the system) do not lead to any emission or absorption of real particles, similar to the persistent tunnelling events in Bloch's case in condensed matter physics.  These tunnelling events simply
   select an appropriate ground state of the system which is a specific superposition of $| k\ra$ states. 
   While each gauge configuration  has definite sign of the topological charge density, the  opposite sign    alternate.  This {\it delicate cancellation} between  the  opposite sign topological configurations  leads to the known result corresponding to   $\la q(x)\ra=0$ in the   physical unperturbed vacuum at $\theta=0$. 
   
     However, when    some external sources  are present in the system  a {\it delicate cancellation} between  the  opposite sign topological configurations  may lead,   in general,  to a local minimum with nontrivial 
   values for  $\la q(x)\ra\neq 0$  in the region $R\sim 10$ fm where external impact is felt, i.e. in the entire volume of collision region. A non-vanishing expectation value for the topological density (which is ${\cal P}$  and  ${\cal CP}$ odd  operator) in large region of space-time can be interpreted, according to eq. (\ref{YM})  as a formation of the   $|\theta_{\rm ind}\ra$ local vacuum state in terminology \cite{Kharzeev:2007tn,Kharzeev:2013ffa}. In fact, one can explicitly show in simplified ``deformed QCD" model that such local minima with nontrivial values of $\la q(x)\ra\neq 0$  indeed exist in      the system, see \cite{Bhoonah:2014gpa} for the details. 
   
   This is precisely the basic idea advocated in refs.\cite{Zhitnitsky:2012im,Zhitnitsky:2012ej} that the long range structure expressed by (\ref{K}),(\ref{YM}) may lead to local violation of the ${\cal P}$  and  ${\cal CP}$ invariance in the same region $R$ where  $\la q(x)\ra\neq 0$ is correlated.     This mechanism is very different from all conventional models (such as sphaleron transitions) which are typically characterized by  a finite correlation length $\sim \Lqcd^{-1}$. Those models   normally  predict a parametrically  small magnitude  $\sim \exp(-\Lqcd R)$ for  coherent  effects considered below, while our mechanism   predicts the intensity   to be proportional $\sim R^{-1}$ where $R\sim 10$ fm    is a size of the system. 
  
  There is a number of generic consequences of this framework   which have been discussed in details   in refs.\cite{Zhitnitsky:2012im,Zhitnitsky:2012ej}. We list them below:
  
   {\bf a)}  The thermal spectrum in $e^+e^-, ~pp$ and $p\bar{p}$ high energy collisions emerges in spite  of the fact that the statistical thermalization could  never be reached in those systems.  An approximate  universality of the temperature   with no dependence on energy of colliding particles nor their nature 
(including $e^+e^-, ~pp$ and $p\bar{p}$ collisions) is due to the fact that the emission occurs from the distorted   QCD vacuum state represented by   the  long range  vacuum    configurations, discussed above,    rather than from the colliding particles  themselves. In heavy ion collisions there is, of course,  a  conventional thermal component, in addition to the emissions due to the tunnelling transitions between  topological $| k\ra$ sectors.

\begin{figure}[t]
 \begin{center}
\includegraphics[width = 0.45\textwidth]{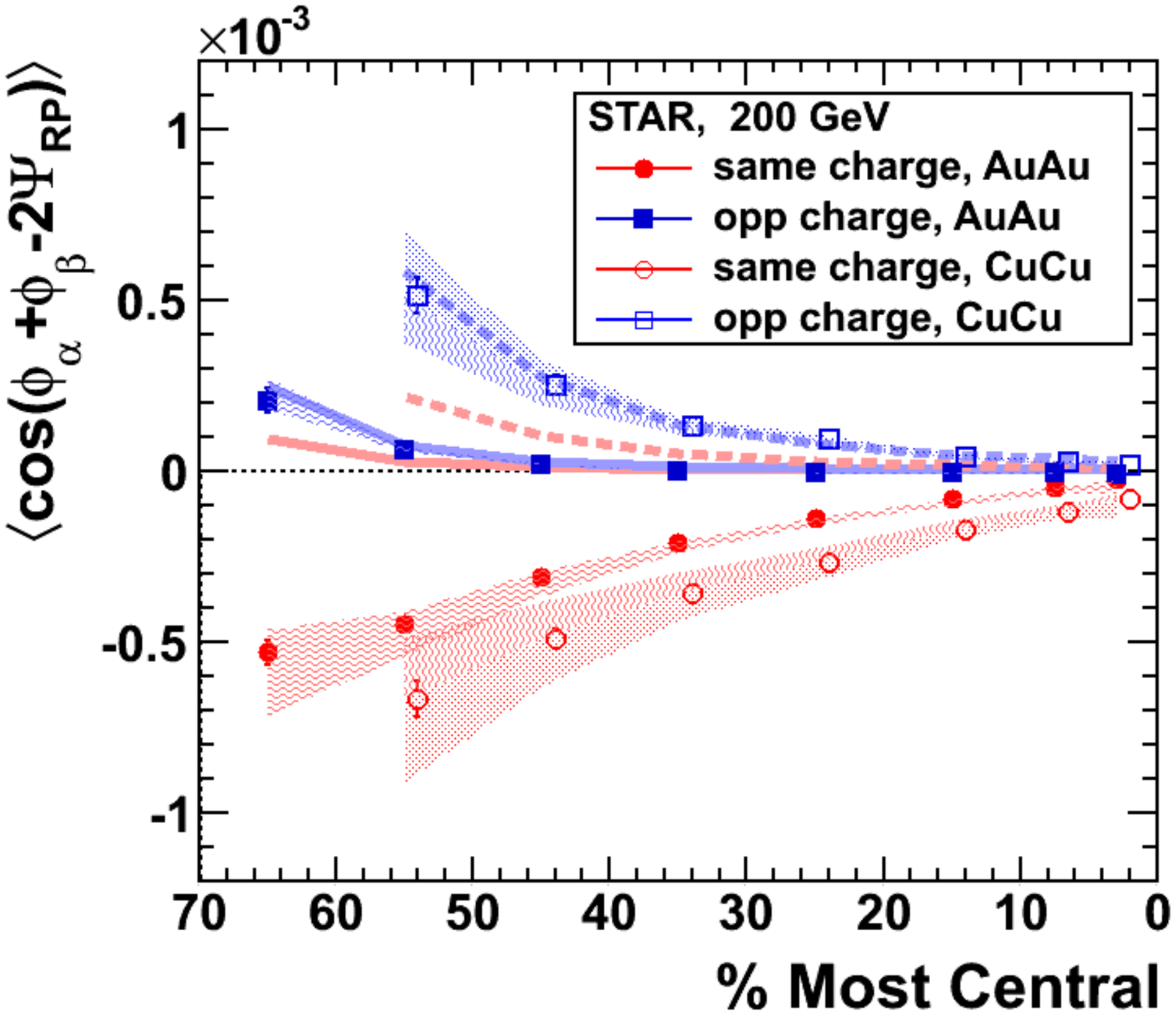}
  \includegraphics[width = 0.45\textwidth]{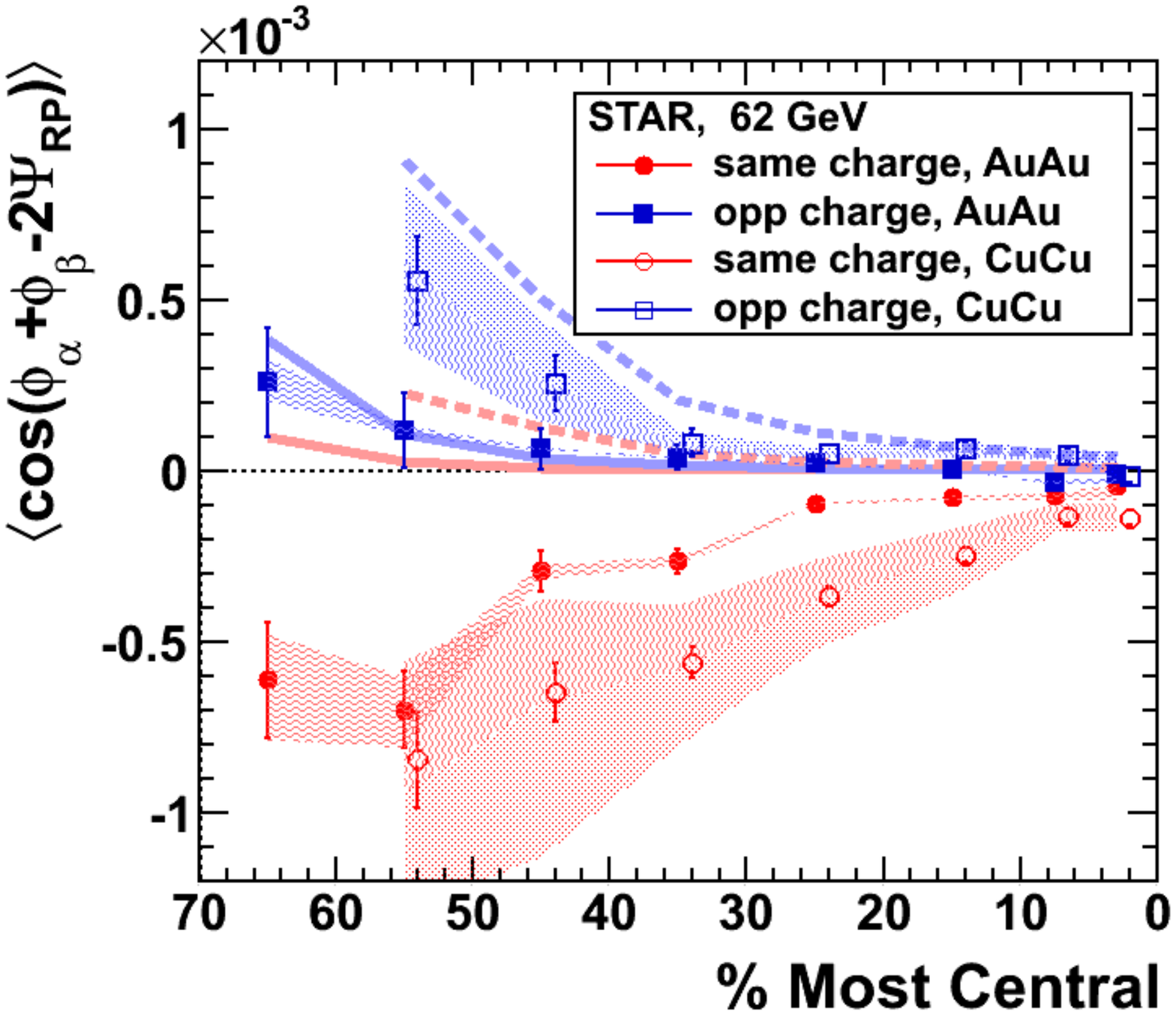}
\caption{\label{RHIC}Data for $\sqrt{s_{NN}}=200\text{GeV}$ and $\sqrt{s_{NN}}=62\text{GeV}$  for Au+Au and Cu+Cu collisions  (adapted  from~\cite{Abelev:2009tx}). 
The plots demonstrate the universality in behaviour, see   text for the details.}     
\end{center}
\end{figure}

{\bf b)} The intensities of the correlations\footnote{The basic observables  which measure corresponding asymmetries on event by event basis have been originally suggested  in \cite{Voloshin:2004vk} and extensively studied in \cite{Abelev:2009tx,Wang:2012qs,Selyuzhenkov:2012py,Abelev:2012pa,Voloshin:2012fv,Selyuzhenkov:2012mf}. We shall not discuss the corresponding definitions   of the observables in the present  work by referring to the original papers.  }   due to the local $\cal{P}$ violation  should demonstrate the universal behaviour similar to the ``universal apparent  thermalization"  discussed in item  {\bf a)}  above  as the source for the both effects  is the same  and related to the emission of the particles due to the distorted tunnelling transitions between  topological $| k\ra$ sectors. In particular, the effect should not depend 
       on energy of colliding ions.
        Such independence on energy is indeed supported by observations        where correlations measured in ${\text{Au+Au }} $ and ${\text{Cu+Cu }} $ collisions at $\sqrt{s_{NN}}= 62~ {\text{GeV}} $ and  $\sqrt{s_{NN}}= 200~ {\text{GeV}} $  
       are almost identical and  independent on energy, see Fig. \ref{RHIC}. We expect the same tendency to continue for the LHC energies. The  recent results from ALICE Collaboration \cite{Selyuzhenkov:2012py,Abelev:2012pa,Voloshin:2012fv,Selyuzhenkov:2012mf} indeed 
       confirm this prediction.

{\bf c)}  For a system of size   $R\gg \Lqcd^{-1}$ the observable  $\cal{P}$ odd  effect due to the collisions   are expected to be  proportional to $R^{-1}$, see \cite{Zhitnitsky:2012im} for the details.  The  $R^{-1}$ scaling essentially describes the deviation of the system from  the  ground state in the region  $R$ as a result of collision.   As one can see from Fig. \ref{RHIC}   some suppression of the measured correlations  with increasing  the size of the system indeed has been observed. We would like to  interpret this suppression as a manifestation of the $R^{-1}$    scaling. 
   Indeed, the effect for ${\text{Au+Au }} $ collisions  with $A\simeq 197$ is obviously suppressed in comparison with 
 ${\text{Cu+Cu }} $ collisions with   $A\simeq 64$. There are   many other effects which influence this ratio (both systems are obviously not very large when derivative expansion used  in \cite{Zhitnitsky:2012im}  is justified). However, the effect goes in the right direction (the effect is stronger for 
  a smaller size system ${\text{Cu+Cu }} $  than for a  larger  ${\text{Au+Au }} $ system).  
  
 Order of magnitude   estimate for the observed    asymmetries     can be presented as follows
    \be
      \label{final}
      \text{[asymmetries  on Figs. \ref{RHIC}]} \sim \frac{e}{N_c^2}\cdot \frac{f}{R\Lqcd}&\sim   10^{-3},  
      \ee
where $e$ is the electric charge $e\sim \sqrt{\alpha}\sim 10^{-1}$,  while  $\sim 1/N_c^2\sim 10^{-1}$  is  a typical suppression  factor which accomponies any phenomena  related to the  $\theta$ dependence, see  \cite{Zhitnitsky:2012im} with a more accurate estimate for this suppression factor.  Finally,  factor $  (R\Lqcd)^{-1}\sim 10^{-1}$ is the key element of the entire framework: all observed asymmetries would be much smaller  $\sim\exp (-R\Lqcd)$  if the long range order (leading to the algebraic $\sim R^{-1}$ decay) is not present in the system. Finally, parameter $f\sim 1$ depends on a number of  different time scales relevant for the problem, see   \cite{Zhitnitsky:2012im} for the details.  

The estimate (\ref{final}) is the direct consequence of  the long range order in QCD formally expressed by eq. (\ref{K}) which explicitly demonstrates  the presence of  topologically protected pole at $k=0$. 
As we discussed in section \ref{contact} the corresponding IR sensitivity is  a result of dynamics  of the topological sectors $|k\ra$  in QCD.   Our application to heavy ion collisions     is based on an idea that the dynamics of the vacuum topological sectors is slightly (but coherently) modified on the scale of order $R\sim (10 ~\rm{fm})$ as a result of a collision (which itself is treated as a time-dependent external impact on the system size $R$).
We have also argued that the corrections to the vacuum energy density due to the long range order   may receive some  power-like corrections of order $1/R$ in contrast with a naive expectation $\exp(-R)$ related to the conventional fluctuations with finite mass $\sim \Lqcd$. As a result of such $1/R$ scaling we estimate   that the intensity for the coherent effects to the local violation of  the $\cal{P}$ and  $\cal{CP}$  invariance  could be quite large (\ref{final}).

To conclude this section: the basic, model-independent consequences of the entire framework listed above are consistent with all available data. Furthermore, the qualitative predictions of this framework (e.g. on energy independence of the  observables at the LHC energies) have been confirmed by the ALICE Collaboration \cite{Selyuzhenkov:2012py,Abelev:2012pa,Voloshin:2012fv,Selyuzhenkov:2012mf}.  Order of magnitude   estimate (\ref{final}) is also consistent with the results from RHIC and the LHC.

\section{Topological Long Range Order  and the de Sitter Accelerating Universe}\label{cosmology}
 
 In this section we consider another application related to the long range order in QCD formally expressed by eq. (\ref{K})   demonstrating  the presence of  topologically protected pole at $k=0$. To be more concrete,  
 we want to consider   external  time-dependent background characterized by the Hubble constant $H\sim 10^{-33}$ eV which is drastically different from the scale $R^{-1}\sim (10 ~\rm{fm})^{-1} $ we dealt with in previous section \ref{ions}. We anticipate some slight   modification of the QCD vacuum topological sectors  $|k\ra$ as a result of slow expansion with rate $H$.    Our goal here is to study the tiny variation of the ground state  energy which may have profound cosmological consequences. 
 
 The logic of our approach   remains the same as in section \ref{ions},
 with the only difference  is that instead of considering time-dependent external impact on the system with size $R$  as a result of ion collisions we are now considering a system with a size of visible Universe $H^{-1}$ in a time dependent background characterized by the Hubble constant $H$. As in  our previous analysis in section \ref{ions}, we try to understand the  qualitative behaviour of the system using a simplified ``deformed  QCD" model, which is weakly coupled gauge theory but preserves all relevant elements,  such as confinement, topological sectors, $\theta$ dependence, etc,   of strongly coupled QCD. 
 
 To be more specific, we address  the following very hard question: How does the non-dispersive  vacuum energy (\ref{YM}), and the corresponding contact term in $\chi$  vary when the system couples to the gravity? We can rephrase and simplify the same question  as follows: how does the rate of tunnelling processes change when the system is considered in a time-dependent background? In principle, the strategy  to carry out  the corresponding computations is as follows. 
  First,  find the classical solution in a nontrivial background which generalize  the fractionally charged monopoles   saturating the contact term (\ref{YM}) along the line of ref.\cite{Thomas:2011ee}. Next, one should  compute the corresponding path integral measure in semiclassical  approximation which  must  depend now on the parameters of a  background such as the Hubble constant $H$. The corresponding corrections (due to the background) in the coefficients of the effective action can be translated to the corresponding corrections in vacuum energy,  which represents the desired result.

   Unfortunately, a complete resolution  of this program is not even feasible. However, there are few general arguments which may provide us with  some hints on possible dependence of the non-dispersive  energy (\ref{YM}) from  a background which will be parameterized in what follows by the Hubble parameter $H$. In general,  one should expect that for sufficiently weak  background   (which we always assume to be the case) the correction to all observables, including the vacuum energy,  can be represented as the power corrections of $H$, i.e.
 \be
 \label{power}
 E_{\rm vac}(H)=\sum_pc_p H^p, ~~~~~~~~~~~~~~~~ c_0=-\frac{N\zeta}{L},
 \ee
 where $c_0$ is given by (\ref{YM}) computed in flat background. 
 
\begin{figure}[t]
 \begin{center}
\includegraphics[width = 0.45\textwidth]{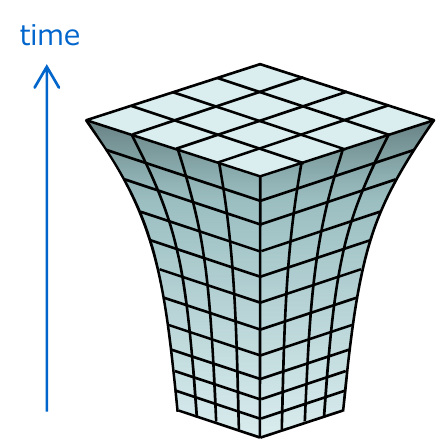}
  \includegraphics[width = 0.45\textwidth]{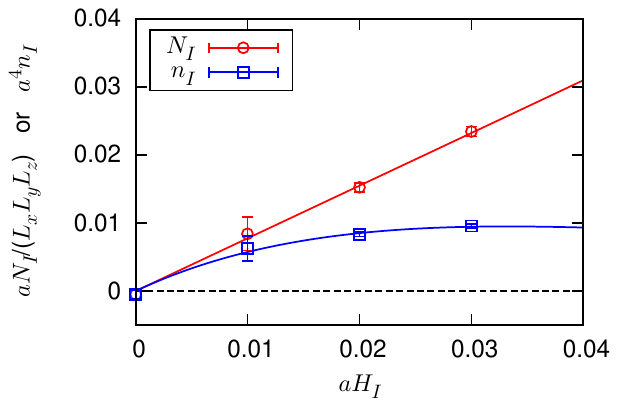}
\caption{\label{H} Expanding lattice (on the left)  and $H$ dependence of the produced particles per unit volume per unit time as  a result of this expansion (on the right). The rate of particle production is shown as a function of the expansion rate $H$ in units of the lattice size $a$. It clearly shows the linear dependence of an observable on $H$, in complete agreement with our eq. (\ref{FLRW}). The plots adapted from  \cite{Yamamoto:2014vda}.}    
\end{center}
\end{figure}

 There is a number  of generic arguments which suggest that $p$ in eq. (\ref{power}) can   only be even, i.e. $p=0, 2, 4...$. The generic arguments, as usual, may have some loopholes.... We refer to ref.\cite{Zhitnitsky:2013pna} with  detail discussions of pros and cons of these arguments. 
 The only comment we want to make here is that an assumption of {\it locality}  supporting $p=0, 2, 4...$ might not be so harmless for non-abelian gauge theories such as QCD, in contrast with a simple example of a massive scalar field theory. Indeed, while QCD has a gap in gauge invariant sectors, it nevertheless demonstrates  a high IR sensitivity in gauge variant sectors in terms of a topologically protected massless pole (\ref{K}). This pole is not screened by the confinement mechanism, and in fact, is the manifestation of the inherently  {\it non-local}    large gauge transformation operator\footnote{\label{T}The large gauge transformation operators is  defined as follows ${\cal{T}}|k\ra=|k+1\ra$. It commutes with the Hamiltonian, $[H, {\cal{T}}]=0$.
The non-dispersive vacuum  energy (\ref{YM}) is non-perturbative in nature as it is generated as a result of tunnelling events between $|k\ra$ and $|k+1\ra$ topological sectors connected by non-local operator $\cal{T}$.} which itself is the key element in the mechanism of generating the non-dispersive vacuum  energy $E_{\rm vac}$.     Furthermore, with few additional simplifications one can explicitly see how the linear $\sim H$ correction may indeed emerge in the ``deformed QCD" model, see Appendix  in ref. \cite{Zhitnitsky:2013pna} with some technical details.    Finally, the linear $\sim H$ correction has been computed  using the  Monte Carlo lattice numerical simulations   in strongly coupled regime \cite{Yamamoto:2014vda}, see Fig.\ref{H} adapted from that paper.

Why is the linear correction $\sim H$ to the vacuum energy  $E_{\rm vac}(H)$  so crucial for our discussions?
A simple answer is that precisely such kind of corrections  will drag   the Universe into de Sitter state as we shall explain  below. 
 \subsection{\label{DE}The de Sitter Accelerating Universe. Dark Energy.}
 In what follows we assume that the linear term $\sim H$ is indeed generated  in the  expression for the vacuum energy in  strongly coupled QCD, such that $ E_{\mathrm{FLRW}}(H)$  in context of the Friedmann-Lema\^itre-Robertson-Walker (FLRW) Universe takes   the  following  form  
\be	\label{FLRW}
  E_{\mathrm{FLRW}}(H)= \left[\Lqcd^4+ H\Lqcd^3+ {\cal{O}}(H^2)\right],
\ee
where the first term $\Lqcd^4$ is well known and well studied expression for the vacuum energy computed in the flat background, 
i.e. $E_{\mathrm{Mink}}=\Lqcd^4$. The presence of a linear term $\sim H$ in this expression is analogous to $1/R$ factor in eq. (\ref{final}), and they    are both generated due to the   long range order present in the system, as discussed in section \ref{contact}. 

As we already explained previously  the energy (\ref{FLRW}) is saturated by tunnelling transitions    between physically identical but topologically distinct topological sectors $|k\ra$.
The physics of tunnelling processes and the corresponding generated energy (\ref{FLRW}) can not be described in terms of any  local dynamical field $\Phi(x)$, as the tunnelling between topologically distinct sectors is fundamentally non-local phenomenon as it is described in terms of non-local operator $\cal{T}$, see footnote 
\ref{T}. 
Therefore, the energy (\ref{FLRW}) can not be expressed in terms of any {\it  local operators such as curvature}, which would be a conventional structure to emerge when physical propagating degrees of freedom are integrated out in the background of the gravitational field.
This feature is similar to the well known property of a topologically ordered phase in condensed matter physics wherein an expectation value of a local operator does not characterize the system.
Instead, a system should be described in terms of some non-local variables such as {\it holonomy}, see 
 \cite{Zhitnitsky:2013pna,Zhitnitsky:2014aja} for the details.
 
Our next assumption, in addition to (\ref{FLRW}),  can be formulated as follows. 
We adopt the paradigm that the relevant definition of the energy  which enters the Einstein equations  in an expanding background (characterized by the parameter $H$)  is the difference $\Delta E (H)\equiv \left[E(H)-E_{\mathrm{Mink}}\right]$, similar to the definition of the Casimir energy.
This element in our analysis is  not very new, and in fact in the present context such a definition for the vacuum energy was advocated for the first time in 1967 by Zeldovich \cite{Zeldovich:1967gd}, see \cite{Sola:2013gha} for review and references on this matter. 

  With these  assumptions just formulated, the Universe has a period of the  de Sitter phase expansion characterized by the
  exponential  growth of the scale  factor  ${\rm a}(t)\sim \exp (H_{0}t)$.
Indeed, the Friedman equation assumes the following form
\be	\label{friedman}
  H^2= \frac{8\pi G}{3}\left(  \rho_{\mathrm{DE}}+\rho_R +\rho_{M}\right)=\frac{8\pi G}{3}\left(   H\Lqcd^3+\rho_R +\rho_{M}\right),  
\ee
where we identify $\rho_{\mathrm{DE}}$ with $\Delta E (H)=\left[E(H) -E_{\mathrm{Mink}}\right]=[E(H) -\Lqcd^4]$ from (\ref{FLRW}).  Furthermore, the corresponding energy density according to eq. (\ref{FLRW}) is given by $\rho_{\mathrm{DE}}= H\Lqcd^3$.
 Two other components:  $\rho_R$ and $\rho_M$ are the radiation and matter (including dark matter) components of the total energy of the system. 
 In (\ref{friedman}) we neglected higher order corrections $ {\cal{O}}(\Lqcd^2 H^2)$ in the expansion (\ref{FLRW}) as $H\ll \Lqcd$.
The radiation and matter components in eq. (\ref{friedman}) scales as $\rho_R\sim {\rm a}^{-4}$ and $\rho_M\sim {\rm a}^{-3}$ correspondingly, such that $\rho_{\mathrm{DE}}$ starts to dominate the universe at some point when $H$ approaches the constant value $H_0$ estimated as follows, which is amazingly close  to the observed value 
\be	\label{H_0}
  H_0\simeq \frac{8\pi G \Lqcd^3 }{3} \simeq  \frac{ \Lqcd^3}{3M^2_{\rm PL}}, ~~  ~~  ~ H_0\sim 10^{-33}~ {\rm eV}, ~~~ \rho_{\mathrm{DE}}\sim H_0\Lqcd^3 \sim \left[(10^{-2}-10^{-3})~ {\rm eV}\right]^4.
\ee
The constant $H_0$, which is unambiguously determined by the strongly coupled dimensional parameter $\Lqcd$ corresponds to the    de Sitter behaviour ${\rm a}(t)\sim \exp (H_{0}t)$ as claimed. 

Few  comments are in order. First, eq. 
(\ref{H_0})  is obviously an   oversimplified  estimate as we do not take into account a number of important elements. In particular, we completely ignore the presence of the light quarks in the system which would lead to factor  $m_q \sim 5$ MeV instead of $\Lqcd\sim 100$ MeV. Furthermore, the computation of the numerical coefficient in front of $\sim H$ in eq. (\ref{FLRW}), which eventually enters the Friedman equation (\ref{friedman})   is a very hard problem. It is obviously the prerogative of  numerical Monte Carlo  lattice simulations similar to computations performed recently in \cite{Yamamoto:2014vda}, rather than a subject of analytical calculations. We think it is next to impossible to perform analytical computations of this coefficient   even in simplified version of QCD, the ``deformed QCD"  as explained at the very beginning of this section. Still, an order of magnitude estimate (\ref{H_0}) looks very promising and very encouraging, and obviously  warrants  a further investigation along this line of thinking. 
 
Our next comment is as follows. 
We advocate a new paradigm which  is based on a fundamentally novel view on the nature and origin of 
the dark energy (DE) corresponding to  the de Sitter  behaviour with constant $H_0$ and constant dark energy
density $\sim H_0\Lqcd^3$ which effectively imitates   the cosmological constant on the time scale when $H_0$ is almost constant (\ref{H_0}). 
This paradigm  is drastically different from the conventional viewpoint that  DE field  is a dynamical local field\footnote{ The model (\ref{friedman}) has been successfully confronted with observations and  it has been claimed that this proposal is consistent with all presently available data, see   references on original works with some comments in \cite{Zhitnitsky:2013pna}. An interesting  unique prediction of this model, yet to be tested,  is that the Universe must demonstrate some sort of the  $\cal{P}$ odd phenomena  on the largest possible scales order of size of the visible Universe $\sim H^{-1}$, similar to our discussions on the local $\cal{P}$ violation in heavy ion collisions in section \ref{ions}. In fact, there are many claims  (from observations) apparently pointing towards a large scale violation of statistical isotropy in our Universe, such as ``cosmic axis of evil", etc. However, I think it is too premature to make any solid statements  on this matter.}.
In our new framework the  DE is   a {\it genuine quantum effect},  and  the DE  effective field is  an auxiliary topological field\footnote{\label{auxiliary}This field does not propagate, does not have a canonical kinetic term, as the sole role of the auxiliary field is to effectively describe the dynamics of the topological sectors of a gauge theory saturating the correlation functions (\ref{top1}), (\ref{K}), (\ref{K1}) in a time-dependent background.  These auxiliary fields are not mandatory fields, but instead play a supplementary role to simplify the analysis of the dynamics of the multiple tunnelling transitions between the distinct topological sectors in strongly coupled QCD.   A similar well-known 
example where the emergent (auxiliary) fields are introduced into the system is the quantum Hall effect. In this case the dynamics of the auxiliary fields are  controlled   by the Chern-Simons effective Lagrangian.}.
The corresponding physics is fundamentally indescribable in terms of any local propagating fields, such as a local scalar field $\Phi(x)$ which is normally used by cosmology- practitioners      to describe  the DE. One should note that a similar auxiliary  field, for example, is known to emerge in the description of the  topologically ordered condensed matter (CM) systems realized in nature.

 \subsection{\label{inflaton}The de Sitter Accelerating Universe. Inflation.}
  It is well known that the deep issue inflation addresses (among many other things) is the origin of the large-scale homogeneity of the observable universe\cite{inflation, linde,mukhanov}. The crucial element of this idea is to have a period of  evolution of the universe which can be 
   well approximated by the de Sitter   behaviour. In this case the scale parameter ${\rm{a}} (t)$ and 
the equation of state takes the following  approximate  form, 
 \be
\label{a}
{\rm{a}}(t)\sim \exp (Ht), ~~~ \rho\approx  -p, ~~~ H\equiv \frac{\dot{\rm{a}}}{\rm{a}}\approx {\rm constant}
\ee
It is commonly  accepted  that such equation of state can be achieved in quantum field theory  by  
  assuming the existence of a local scalar matter field  $\Phi (x) $, the {\it inflaton},  with a non-vanishing potential energy density $V[\Phi (x)]$. The shape of this potential energy can be adjusted in a such a way that the  contribution to energy density $\rho$ and pressure $p$  is in agreement with the above equation of state.

We advocate a fundamentally different view on the nature and origin of the inflaton field \cite{Zhitnitsky:2013pna,Zhitnitsky:2014aja}. Our proposal is based on assumption that the inflaton is an auxiliary field of a strongly coupled gauge theory, very much the same as we discussed in previous section with application to  the dark energy (DE), see footnote \ref{auxiliary} with some comments.    However, the scale of inflation and scale of the present DE density (\ref{H_0}) are drastically different. Therefore, we assume there existence of a scaled up version of QCD (which is coined in ref. \cite{Zhitnitsky:2013pna} as $\qcd$) determined by high inflationary scale  $\Lbar< M_{PL}$, to be determined from observations.  A similar construction (though in a different context) had been suggested long ago for a different purpose and is known as technicolor. 

Therefore, one can repeat the arguments from the previous section \ref{DE} to conclude  that   the Friedman equation has a non-trivial solution $H_0\simeq $ constant,  
\be	\label{friedman-infl}
  H^2= \frac{8\pi G}{3}\left(  \rho_{\mathrm{Inf}}+\rho_R\right)=\frac{8\pi G}{3}\left(H\Lbar^3+\rho_R\right),  
  ~~ \Rightarrow ~~~H_0\simeq \frac{8\pi G  \Lbar^3}{3}, ~~~ \rho_{\mathrm{Inf}}\simeq  \frac{8\pi G  \Lbar^6}{3}. ~~~
  \ee
This solution $H_0\simeq $ constant  is identified with   inflationary de Sitter behaviour   (\ref{a}).  
In our new framework the inflation is   a  genuine quantum effect while  the {\it  inflaton}  is    an auxiliary topological field, see footnote \ref{auxiliary} with few comments on the nature of this field. 

The inflationary regime described by eqs. (\ref{a}) and  (\ref{friedman-infl}) would be the final destination of our Universe if the interaction of the $\qcd$ fields with  the Standard Model (SM)  particles were always switched off.
When the coupling is switched back on, the end of inflation is triggered precisely by this interaction which itself is unambiguously fixed by the triangle anomaly as we discuss below. This is  so-called the reheating period.
  The only information which is needed for the future discussions is that the auxiliary topological field $b(x)$,   generating the corresponding inflaton-related energy  (\ref{friedman-infl}) couples to the SM particles precisely in the same way as the  $\theta$ parameter couples  to the gauge fields, see \cite{Zhitnitsky:2013pna} for the details. 
In other words, the coupling is  
\be	\label{coup}
  {\cal L}_{b\gamma\gamma} (x)= \frac{\alpha (H_0)}{8\pi} N  Q^2 \left[ \theta- b(x)\right] \cdot F_{\mu\nu} \tilde F^{\mu\nu} (x) \, ,
\ee
where $\alpha(H_0)$ is the fine-structure constant measured during the period of inflation, $Q$ is the electric charge of a $\qcd$ quark, $N$ is the number of colours of the strongly coupled $\qcd$, and $F_{\mu\nu}$ is the usual electromagnetic field strength.
  The coupling  of the $b(x)$ with other $E\&W$ gauge bosons can be unambiguously reconstructed as explained in \cite{Zhitnitsky:2013pna}, but we keep a single $E\&M$ field $F_{\mu\nu}$ to simplify the notations and emphasize on the crucial elements of this coupling which is unambiguously  fixed by quantum anomaly. We take $\theta=0$ in eq. (\ref{coup}) as we do not intend to discuss in this work an interesting, but different, subject related to the dynamics of the physical axion field.  
  
  As we explained  above the  $b(x)$-field (which is not dynamical, but rather, an emergent auxiliary field, see footnote \ref{auxiliary} with few comments)   should be treated as a coherent field representing the rate of tunnelling events in the system. It  varies and  fluctuates as a consequence  of expansion, rather than a result of the presence of a kinetic term.  As a result of these fluctuations in time dependent background $b(x)$ field radiates real physical particles in expanding universe.  This radiation occurs  in spite  of the fact that $b(x)$ itself is not a dynamical field. This is precisely the way how the inflaton energy (\ref{friedman-infl})   can be transferred   to   the SM particles.  Eventually this process of the energy transfer is   responsible for  the termination of the inflationary epoch.

How long does it take for the energy (\ref{friedman-infl}) to be transferred to SM particles? 
 The corresponding rate of transfer is determined by the coupling (\ref{coup}).
 We estimate  the production of real particles in expanding background parameterized by the Hubble constant $H_0$ as follows 
 \be	\label{production}
  \frac{dP}{dV dt}\sim \alpha^2(H_0) H_0\Lbar^3.
\ee
The particle production in our system is a result of multiple tunnelling events between the topological sectors $|k\ra$ in the background of the gravitational field. It should be contrasted with conventional computations during the reheating epoch when the   radiation of SM particles occur as a result of dynamics of real propagating inflaton field.  As we emphasized previously, the {\it inflaton} in our framework is not a propagating degree of freedom, but rather is an auxiliary, non-propagating topological field, see footnote \ref{auxiliary} with few comments. 

 The crucial element in  formula (\ref{production}) is the linear dependence on $H$ which is precisely the same effect as we discussed after  eq. (\ref{FLRW}). The difference between  (\ref{FLRW})  and  (\ref{production})  is that   eq. (\ref{FLRW}) describes the real part of energy momentum tensor, while formula (\ref{production}) represents the imaginary (absorptive) portion  of the energy momentum tensor describing the production of real particles. However, the  analyticity suggests that if such term is present in (\ref{FLRW}) it must be also present in  (\ref{production}), and vice versa. The lattice numerical Monte Carlo simulations also confirm 
the linear dependence on $H$ for  particle production in strongly coupled QCD in the de Sitter background \cite{Yamamoto:2014vda}. This computation is  consistent    with our formula (\ref{production}) and, therefore, it  strongly supports our basic assumption on    presence of the linear correction $\sim H$  to the vacuum energy (\ref{FLRW}).  

  At the moment $\tau_{\rm Inf}$ the inflation comes to the end  as the dominant portion of the energy is   transferred to the light SM particles.
Based on particle production rate (\ref{production}) one could estimate 
the number of $e$-folds  in this mechanism when this moment arrives \cite{Zhitnitsky:2013pna,Zhitnitsky:2014aja},
\be	\label{e-folds}
  \tau_{\rm Inf}^{-1}\sim  {H_0\alpha^2(H_0)}, ~~~~\Longrightarrow ~~~N_{\text{Inf}}\sim \frac{1}{\alpha^2(H_0)},
\ee   
which leads to an estimate $N_{\text{Inf}}\simeq 100$ for $\alpha(H_0)\sim 0.1$, see \cite{Zhitnitsky:2014aja}
with more accurate numerical estimate. 
The key element of this $\qcd$ inflationary scenario is that the number of $e$-folds $N_{\text{Inf}}$  in this framework is determined by the gauge coupling constant $\alpha(H_0)$ rather than by dynamics of an ad hoc inflaton $\Phi(x)$ field  governed by an ad hoc  inflationary potential $V[\Phi (x)]$.

\section{Conclusion}\label{conclusion}
In the present work we advocate an idea that 
QCD has some ``hidden" long range order which was suspected long ago  \cite{Luscher:1978rn}. 
 It may have  profound consequences for our understanding of gauge theories if this long range order will be confirmed by future numerical and analytical computations. In particular, we have argued that the observation of the local violation of $\cal{P}$ invariance in heavy ion collisions as discussed in section \ref{ions} might be a simple    consequence of this long range order. We argue that a  sufficiently high intensity of the effect (\ref{final}) is a result of a coherent modification of the ground state on  the  scale of  nuclei with size $R\sim 10$ fm, rather than a result of very  strong $\cal{P}$  odd  fluctuation  with typical QCD size of 1 fm. 
 
 Our applications to cosmology considered in section \ref{DE} may have  some  profound consequences
 for our understanding of the Universe we live in.  In particular, we argued that the de Sitter behaviour might be dynamically generated as a result of the long range order advocated in this work. 
 In this framework, the  {\it DE}  field is   not  fundamental local field. Instead,  the  {\it DE} field    
   is  an auxiliary   topological field which effectively describes the dynamics of topological sectors in QCD   when it is considered in the expanding universe.  The corresponding energy in this framework has fundamentally different nature than conventional energy  when a theory is formulated  in terms of a local  dynamical  field $\Phi (x) $, for example in the Higgs model. In particular, it can not be expressed in terms of any propagating physical degrees of freedom as the corresponding energy  has a non-dispersive nature. It also can not be expressed in terms of any local variables, as this phenomenon is  inherently non-local. The same arguments also apply to the {\it inflaton}, see section \ref{inflaton}. 
   
We conclude this work (mainly devoted to the long range order in QCD and some profound cosmological consequences of this long range order) with the following comment about a different field of physics with drastically different scales.
Namely, as we discussed at length in this paper, the  heart of the proposal is a fundamentally new type of energy (\ref{YM}), (\ref{FLRW}) which is not related to  any propagating degrees of freedom. 
Rather, this novel (non-dispersive) contribution to the energy has genuine quantum nature. The effect is formulated in terms of the tunnelling processes between topologically different but physically identical states.
This novel type of energy, in fact, has been well studied in the QCD lattice simulations in the flat background, see  \cite{Zhitnitsky:2013pna} for references on the original works.
Our comment relevant for the present study is that this fundamentally new type of energy can be, in principle, studied in a tabletop experiment by measuring some specific corrections to the Casimir vacuum energy  in the Maxwell theory as suggested in  \cite{Cao:2013na,Zhitnitsky:2013hba,Zhitnitsky:2014dra}.
This fundamentally new contribution to the Casimir pressure emerges as a result of tunnelling processes, rather than due to the conventional fluctuations of the propagating photons with two physical polarizations.
This effect does not occur for the scalar field theory, in contrast with conventional Casimir effect which is operational for both: scalar as well as for Maxwell fields.
The extra energy computed in \cite{Cao:2013na,Zhitnitsky:2013hba,Zhitnitsky:2014dra} is the direct analog of the non-dispersive contribution to the energy  (\ref{YM}), (\ref{FLRW})   which is the key player of the present work.
In fact, an extra contribution to the Casimir pressure emerges in this system as a result of nontrivial holonomy which can be enforced by  the nontrivial boundary conditions imposed in ref \cite{Cao:2013na,Zhitnitsky:2013hba,Zhitnitsky:2014dra}.  

 \section*{Acknowledgements}
I am thankful to Larry McLerran,  Slava Mukhanov, Nemanja Kaloper,  and  other participants of the  International Conference on New Frontiers in Physics (ICNFP), Crete, August 2014, where this work was presented. 
This research was supported in part by the Natural Sciences and Engineering Research Council of Canada.


\begin{thebibliography}{99}

\bibitem{Voloshin:2004vk}
  S.~A.~Voloshin,
  Phys.\ Rev.\  C {\bf 70}, 057901 (2004)
  [arXiv:hep-ph/0406311].


 \bibitem{Abelev:2009tx}
  B.~I.~Abelev {\it et al.}  [STAR Collaboration],
  Phys.\ Rev.\  C {\bf 81}, 054908 (2010)
  [arXiv:0909.1717 [nucl-ex]].
  
\bibitem{Wang:2012qs} 
  G.~Wang [STAR Collaboration],
  proceedings for Quark Matter 2012,
  arXiv:1210.5498 [nucl-ex].

\bibitem{Selyuzhenkov:2012py} 
  I.~Selyuzhenkov [ALICE Collaboration],
  PoS WPCF {\bf 2011}, 044 (2011)
  [arXiv:1203.5230 [nucl-ex]].
  
\bibitem{Abelev:2012pa} 
  B.~Abelev {\it et al.}  [ALICE Collaboration],
  Phys. \ Rev. \ Lett. {\bf 110}, 012301 (2013)
  arXiv:1207.0900 [nucl-ex].
   
\bibitem{Voloshin:2012fv} 
  S.~A.~Voloshin [ALICE Collaboration],
  Proceedings of the Quark Matter 2012 Conference, Washington D.C., August 2012, 
  arXiv:1211.5680 [nucl-ex].
  
\bibitem{Selyuzhenkov:2012mf} 
  I.~Selyuzhenkov,
  FAIRNESS 2012 workshop, 3-8 September 2012, Hersonissos, Greece, 
    arXiv:1212.5489 [nucl-ex].

\bibitem{Kharzeev:2007tn}
  D.~Kharzeev and A.~Zhitnitsky,
  Nucl.\ Phys.\  A {\bf 797}, 67 (2007)
  [arXiv:0706.1026 [hep-ph]].


\bibitem{Kharzeev:2013ffa} 
  D.~E.~Kharzeev,
  Prog.\ Part.\ Nucl.\ Phys.\  {\bf 75}, 133 (2014)
  [arXiv:1312.3348 [hep-ph]].
  
\bibitem{Kharzeev:2007jp}
  D.~E.~Kharzeev, L.~D.~McLerran and H.~J.~Warringa,
  Nucl.\ Phys.\  A {\bf 803}, 227 (2008)
  [arXiv:0711.0950 [hep-ph]].

\bibitem{Fukushima:2008xe}
  K.~Fukushima, D.~E.~Kharzeev and H.~J.~Warringa,
  Phys.\ Rev.\  D {\bf 78}, 074033 (2008)
  [arXiv:0808.3382 [hep-ph]].
  
\bibitem{Zhitnitsky:2013hs} 
  A.~R.~Zhitnitsky,
  Annals Phys.\  {\bf 336}, 462 (2013)
  [arXiv:1301.7072 [hep-ph]].
  
    
    
  
\bibitem{Zhitnitsky:2012im} 
  A.~R.~Zhitnitsky,
  Nucl.\ Phys.\ A {\bf 886}, 17 (2012)
  [arXiv:1201.2665 [hep-ph]].
  
\bibitem{Zhitnitsky:2012ej} 
  A.~R.~Zhitnitsky,
  Nucl.\ Phys.\ A {\bf 897}, 93 (2013)
  [arXiv:1208.2697 [hep-ph]].
  
\bibitem{Zhitnitsky:2013pna} 
  A.~R.~Zhitnitsky,
  Phys.\ Rev.\ D {\bf 89}, 063529 (2014)
  [arXiv:1310.2258 [hep-th]].
  
\bibitem{Zhitnitsky:2014aja} 
  A.~R.~Zhitnitsky,
  Phys.\ Rev.\ D {\bf 90}, 043504 (2014)
  [arXiv:1404.5965 [hep-ph]].
  
  \bibitem{witten}
  E.~Witten,
  Nucl.\ Phys.\  B {\bf 156}, 269 (1979).

 \bibitem{ven}
G.~Veneziano,
  Nucl.\ Phys.\  B {\bf 159}, 213 (1979).

\bibitem{vendiv}
  P.~Di Vecchia and G.~Veneziano,
  Nucl.\ Phys.\  B {\bf 171}, 253 (1980).
  
\bibitem{Ilgenfritz:2007xu} 
  E.~-M.~Ilgenfritz, K.~Koller, Y.~Koma, G.~Schierholz, T.~Streuer and V.~Weinberg,
  Phys.\ Rev.\ D {\bf 76}, 034506 (2007)
  [arXiv:0705.0018 [hep-lat]].
  

 
\bibitem{Ilgenfritz:2008ia} 
  E.~-M.~Ilgenfritz, D.~Leinweber, P.~Moran, K.~Koller, G.~Schierholz and V.~Weinberg,
  Phys.\ Rev.\ D {\bf 77}, 074502 (2008)
  [Erratum-ibid.\ D {\bf 77}, 099902 (2008)]
  [arXiv:0801.1725 [hep-lat]].
  
\bibitem{Bruckmann:2011ve} 
  F.~Bruckmann, F.~Gruber, N.~Cundy, A.~Schafer and T.~Lippert,
  Phys.\ Lett.\ B {\bf 707}, 278 (2012)
  [arXiv:1107.0897 [hep-lat]].


\bibitem{Luscher:1978rn} 
  M.~Luscher,
  Phys.\ Lett.\ B {\bf 78}, 465 (1978).

\bibitem {Yaffe:2008}
  M.~\"{U}nsal and L.~G.~Yaffe,
  Phys.\ Rev.\ D {\bf 78}, 065035 (2008).
  [arXiv:0803.0344 [hep-th]].


\bibitem{Thomas:2011ee} 
  E.~Thomas and A.~R.~Zhitnitsky,
  Phys.\ Rev.\ D {\bf 85}, 044039 (2012)
  [arXiv:1109.2608 [hep-th]].
  
\bibitem{Bhoonah:2014gpa} 
  A.~Bhoonah, E.~Thomas and A.~R.~Zhitnitsky,
  Nucl.\ Phys.\  B {\bf xxx},  to appear. 
  [arXiv:1407.5121 [hep-ph]].

\bibitem{Yamamoto:2014vda} 
  A.~Yamamoto,
  Phys.\ Rev.\ D {\bf 90}, 054510 (2014)
  [arXiv:1405.6665 [hep-lat]].
  
\bibitem{Zeldovich:1967gd}
  Y.~B.~Zeldovich,
  JETP Lett.\  {\bf 6}, 316 (1967)
  [Pisma Zh.\ Eksp.\ Teor.\ Fiz.\  {\bf 6}, 883 (1967)].



\bibitem{Sola:2013gha} 
  J.~Sola,
  J.\ Phys.\ Conf.\ Ser.\  {\bf 453}, 012015 (2013)
  [arXiv:1306.1527 [gr-qc]].

\bibitem{inflation} A. Guth, Phys. Rev.  {\bf  D 23} (1981) 347;\\ A. Linde, Phys. Lett.  {\bf B 108} (1982) 389.
 

 
\bibitem{linde}
  A. D. Linde, Inflationary Cosmology, Lect. Notes Phys. {\bf 738}, 1 (2008) [arXiv:0705.0164 [hep-th]]
  
   \bibitem{mukhanov}
V.~Mukhanov, {\it Physical Foundation of Cosmology}, Cambridge Univ.\ Pr.\ , 2005.
 

    
\bibitem{Cao:2013na} 
  C.~Cao, M.~van Caspel and A.~R.~Zhitnitsky,
  Phys.\ Rev.\ D {\bf 87}, no. 10, 105012 (2013)
  [arXiv:1301.1706 [hep-th]].
 
\bibitem{Zhitnitsky:2013hba} 
  A.~R.~Zhitnitsky,
  Phys.\ Rev.\ D {\bf 88}, no. 10, 105029 (2013)
  [arXiv:1308.1960 [hep-th]].
 
\bibitem{Zhitnitsky:2014dra} 
  A.~Zhitnitsky,
  Phys.\ Rev.\ D {\bf 90}, no. 10,  105007 (2014)
  [arXiv:1407.3804 [hep-th]].
       
\end{thebibliography}
\end{document}